\documentclass[twocolumn,showpacs,preprintnumbers,pra]{revtex4}
\usepackage{amssymb}
\usepackage{amsmath}
\usepackage{graphicx}
\usepackage{dcolumn}
\usepackage{bm}

\begin{document}

\title{Exactly Measurable Concurrence of Mixed States}
\author{Chang-shui Yu}
\author{C. Li}
\author{He-shan Song}
\email{hssong@dlut.edu.cn}
\affiliation{School of Physics and Optoelectronic Technology, Dalian University of
Technology, Dalian 116024, P. R. China}
\date{\today }

\begin{abstract}
We, for the first time, show that bipartite concurrence for rank 2
mixed states of qubits is written by an observable which can be
exactly and directly measurable in experiment by local projective
measurements, provided that four copies of the composite quantum
system are available. In addition, for a tripartite quantum pure
state of qubits, the 3-tangle is also shown to be measurable only by
projective measurements on the reduced density matrices of a pair of
qubits conditioned on four copies of the state.
\end{abstract}

\maketitle

\section{Introduction}

Quantum entanglement has been realized to be a useful physical resource for
quantum information processing. The quantification of entanglement has
attracted increasing interests in recent years [1]. However, entanglement
\textit{per se} is not an observable in strict quantum mechanical sense.
That is to say, up till now, no directly measurable observable corresponds
to entanglement of a given arbitrary quantum state, owing to the unphysical
quantum operations in usual entanglement measure [2], for example, the
complex conjugation of concurrence [3] and the partial transpose of
negativity [4,5]. A general approach to characterize entanglement in
experiment is quantum state tomography, by which one needs to first
reconstruct the density matrix by measuring a complete set of observables
[6-8] and then turn to the mathematical problem of evaluating some
entanglement measure. But this approach has been implemented successfully in
experiment which is only suitable for small quantum systems because the
number of measured observables grows rapidly with the dimension of the
system. Entanglement witnesses are also effective for the detection of
entanglement [9] if priori knowledge on the states is available, which
depends on the detected states. Quite recently, some approaches have been
reported for the determination of entanglement in experiment
[2,10-12,14-16]. The most remarkable are the new formulation of concurrence
[13] in terms of copies of the state which led to the first direct
experimental evaluation of entanglement [12] and some analogous
contributions [14] to multipartite concurrence. These are only restrictive
to pure states. However, in reality quantum states are never really pure,
hence it is very necessary to consider how to determine the entanglement of
a mixed state in experiment.

In this paper, we study the exact and measurable entanglement of
mixed states. Recently concurrence in terms of copies of states was
generalized to bipartite [15] and multipartite mixed states [16]
such that concurrence of mixed states can be obtained by effective
experimental estimations. However, both the two estimations in fact
only provide observable lower bounds of concurrence with purity
dependent tightness for mixed states instead of the exact
concurrence. One has to accept the fact that the exactly measurable
entanglement for a general mixed state is still a challenge. Here we
only take the first step to focus on concurrence for rank 2
bipartite mixed states of qubits. The exact value instead of lower
bound is given by a Hermite operator (observable) and shown to be
directly measurable in experiment based on local projective
measurements, provided that four copies of the tripartite quantum
pure state are available. It is quite interesting that, for a given
tripartite quantum pure state of qubits, its 3-way entanglement
measure i.e., 3-tangle can be directly measured only by the local
projective measurements on the reduced density matrices of any two
qubits without the requirement. of performing any operation on the
third qubit.

The paper is organized as follows. In Sec. II, we show that the exact
concurrence is directly and locally measurable in experiment, and that the
exact 3-tangle for a tripartite quantum pure state of qubits is also
measurable conditioned on fourfold copy, respectively. The conclusion is
drawn in Sec. III.

\section{Measurable Concurrence}

We first show that the concurrence of a rank-2 mixed state is exactly
measurable. The concurrence of a given rank 2 mixed state of a pair of
qubits $\rho _{AB}$ is given [3,17] by
\begin{equation}
C\left( \rho _{AB}\right) =\left\vert \lambda _{1}-\lambda _{2}\right\vert ,
\end{equation}%
where $\lambda _{i}$ are the square roots of the eigenvalues of $\rho _{AB}%
\tilde{\rho}_{AB}$ in decreasing order with $\tilde{\rho}_{AB}=\left( \sigma
_{y}\otimes \sigma _{y}\right) \rho _{AB}^{\ast }\left( \sigma _{y}\otimes
\sigma _{y}\right) $. After a simple algebra, one has%
\begin{equation}
C^{2}\left( \rho _{AB}\right) =Tr\left( \rho _{AB}\tilde{\rho}_{AB}\right)
-\tau ,
\end{equation}%
where
\begin{equation}
\tau =\sqrt{2\left\{ \left[ Tr\left( \rho _{AB}\tilde{\rho}_{AB}\right) %
\right] ^{2}-Tr\left[ \left( \rho _{AB}\tilde{\rho}_{AB}\right) ^{2}\right]
\right\} }
\end{equation}%
is half of the 3-tangle [18,19] of the tripartite quantum pure state of
qubits such that $\rho _{AB}$ is the reduced density matrix. It is easy find
that
\begin{equation}
Tr\left( \rho _{AB}\tilde{\rho}_{AB}\right) =\sqrt{Tr[(\rho _{AB}\otimes
\rho _{AB})\mathcal{B}]},
\end{equation}%
with $\mathcal{B}=4P_{-}^{A_{1}A_{2}}\otimes P_{-}^{B_{1}B_{2}}$, where%
\begin{equation}
P_{-}^{(i_{m}i_{n})}=\frac{1}{\sqrt{2}}\left( \left\vert 0\right\rangle
_{i_{m}}\left\vert 1\right\rangle _{i_{n}}-\left\vert 1\right\rangle
_{i_{m}}\left\vert 0\right\rangle _{i_{n}}\right) ,
\end{equation}%
denoting the projector onto the anti-symmetric subspace $\mathcal{H}%
_{i_{m}}\wedge \mathcal{H}_{i_{n}}$ of $\mathcal{H}_{i_{m}}\otimes \mathcal{H%
}_{i_{n}}$ where $i=A,B$ corresponds to the subsystems, and $m,n=1,2,3,4$
marks the different copies of $\rho _{AB}$. $Tr\left( \rho _{AB}\tilde{\rho}%
_{AB}\right) $ has been written in the form of the expectation value of the
self-adjoint operator $\mathcal{B}$ which is measured by local projective
measurements on two copies of $\rho _{AB}$, hence it is measurable. The
remaining are to show that $Tr\left[ \left( \rho _{AB}\tilde{\rho}%
_{AB}\right) ^{2}\right] $ is also directly measurable in experiment.

Consider a decomposition of
\begin{equation}
\rho _{AB}=\sum\limits_{i}\left\vert \psi _{i}\right\rangle
_{AB}\left\langle \psi _{i}\right\vert =\sum\limits_{i}p_{i}\left\vert
\varphi _{i}\right\rangle _{AB}\left\langle \varphi _{i}\right\vert ,
\end{equation}%
with $\sum\limits_{i}p_{i}=1$ and $\left\vert \psi _{i}\right\rangle _{AB}=%
\sqrt{p_{i}}\left\vert \varphi _{i}\right\rangle _{AB}$, $Tr\left[ \left(
\rho _{AB}\tilde{\rho}_{AB}\right) ^{2}\right] $ can be given by
\begin{eqnarray}
&&Tr\left[ \left( \rho _{AB}\tilde{\rho}_{AB}\right) ^{2}\right]   \notag \\
&=&Tr\sum\limits_{ijkl}\left\vert \psi _{i}\right\rangle ^{1}\left\langle
\psi _{i}\right\vert \Sigma \left\vert \psi _{j}^{\ast }\right\rangle
^{2}\left\langle \psi _{j}^{\ast }\right\vert \Sigma \left\vert \psi
_{k}\right\rangle ^{3}\left\langle \psi _{k}\right\vert \Sigma \left\vert
\psi _{l}^{\ast }\right\rangle ^{4}\left\langle \psi _{l}^{\ast }\right\vert
\Sigma   \notag \\
&=&\sum\limits_{ijkl}\left\langle \psi _{i}\right\vert ^{1}\left\langle \psi
_{j}\right\vert ^{2}\mathbf{P}\left\vert \psi _{j}\right\rangle
^{2}\left\vert \psi _{k}\right\rangle ^{3}\left\langle \psi _{k}\right\vert
^{3}\left\langle \psi _{l}\right\vert ^{4}\mathbf{P}\left\vert \psi
_{l}\right\rangle ^{4}\left\vert \psi _{i}\right\rangle ^{1}  \notag \\
&=&\sum\limits_{ijkl}\left\langle \psi _{i}\right\vert ^{1}\left\langle \psi
_{j}\right\vert ^{2}\left\langle \psi _{k}\right\vert ^{3}\left\langle \psi
_{l}\right\vert ^{4}\left( \mathbf{P\otimes P}\right) \left\vert \psi
_{j}\right\rangle ^{2}\left\vert \psi _{k}\right\rangle ^{3}\left\vert \psi
_{l}\right\rangle ^{4}\left\vert \psi _{i}\right\rangle ^{1}  \notag \\
&=&Tr\left[ \otimes _{i=1}^{4}\rho _{AB}^{i}\left( \mathbf{P\otimes P}%
\right) SWAP\right] =Tr\left[ \otimes _{i=1}^{4}\rho _{AB}^{i}\mathcal{A}%
\right] ,
\end{eqnarray}%
with
\begin{equation}
\mathcal{A}=\frac{\left( \mathbf{P\otimes P}\right) SWAP+SWAP^{\dagger
}\left( \mathbf{P\otimes P}\right) }{2},
\end{equation}%
\begin{equation}
\mathbf{P=}P_{-}\otimes P_{-}
\end{equation}%
where $P_{-}$ without superscripts and subscripts means a general projector
onto the anti-symmetric subspace of two qubits, $\Sigma =\sigma _{y}\otimes
\sigma _{y}$, $SWAP$ is defined as%
\begin{equation*}
\left\vert \psi _{j}\right\rangle ^{2}\left\vert \psi _{k}\right\rangle
^{3}\left\vert \psi _{l}\right\rangle ^{4}\left\vert \psi _{i}\right\rangle
^{1}=SWAP\left\vert \psi _{i}\right\rangle ^{1}\left\vert \psi
_{j}\right\rangle ^{2}\left\vert \psi _{k}\right\rangle ^{3}\left\vert \psi
_{l}\right\rangle ^{4},
\end{equation*}%
and all $\left\vert \psi \right\rangle $ denote bipartite pure state with
subscirpts \textit{AB} omitted. The superscripts of $\left\vert \psi
\right\rangle $ in eq. (7) mark the different copies of $\rho _{AB}$. The
last "=" in eq. (7) follows from the fact that $Tr\left[ \left( \rho _{AB}%
\tilde{\rho}_{AB}\right) ^{2}\right] $ should not be changed if the
different copies are exchanged. It happened that $\mathcal{A}$ can always be
written by%
\begin{equation*}
\mathcal{A=}\frac{1}{2}\left( M_{A}\otimes M_{B}-N_{A}\otimes N_{B}\right) ,
\end{equation*}%
where $M_{A}=M_{B}$ and $N_{A}=N_{B}$, $M_{i}$ and $N_{i}$ denote the
observables (Hermite operators) of the four copies of the \textit{i}th
subsystem. In principle, one can always obtain the projectors of $M_{i}$ and
$N_{i}$ by their eigenvalue decompositions and perform corresponding local
projective measurements on the \textit{fourfold} subsystems. We have found
that both $M_{i}$ and $N_{i}$ are rank 2, hence $\mathcal{A}$ can be
directly measured by 8 projectors. That is to say, the exact concurrence can
be directly measured by \textit{10} local projective measurements (plus $%
\mathcal{B}$ ).

In fact one can reduce the number of projective measurements further. After
a simple derivation, one can find that $M_{i}$ can always be written by%
\begin{equation}
M_{i}=\frac{\sqrt{2}}{2}\left( P_{-}^{i_{1}i_{2}}\otimes
P_{-}^{i_{3}i_{4}}-P_{-}^{i_{1}i_{3}}\otimes
P_{-}^{i_{2}i_{4}}-P_{-}^{i_{1}i_{4}}\otimes P_{-}^{i_{2}i_{3}}\right) .
\end{equation}%
Thus $M_{A}\otimes M_{B}$ corresponds to \textit{9} groups of projective
measurements (\textit{9} observables). Consider the invariance of the
exchange of different copies, one can find that $M_{A}\otimes M_{B}$ can be
reduced to \textit{2} groups of projective measurements as%
\begin{eqnarray}
M_{A}\otimes M_{B} &=&\frac{1}{2}\left( 3P_{-}^{A_{1}A_{2}}\otimes
P_{-}^{A_{3}A_{4}}\otimes P_{-}^{B_{1}B_{2}}\otimes
P_{-}^{B_{3}B_{4}}\right.   \notag \\
&&\left. -2P_{-}^{A_{1}A_{2}}\otimes P_{-}^{A_{3}A_{4}}\otimes
P_{-}^{B_{1}B_{3}}\otimes P_{-}^{B_{2}B_{4}}\right) .
\end{eqnarray}%
$Tr\left( \rho _{AB}\tilde{\rho}_{AB}\right) $ is also included in the two
measurements. $N_{i}$ can be explicitly by%
\begin{equation}
N_{i}=\sqrt{3}\left( \left\vert \Psi \right\rangle \left\langle \Psi
\right\vert -\left\vert \bar{\Psi}\right\rangle \left\langle \bar{\Psi}%
\right\vert \right) ,
\end{equation}%
where
\begin{eqnarray}
\left\vert \Psi \right\rangle  &=&\frac{1}{\sqrt{6}}\left( \left\vert
0011\right\rangle +e^{\frac{2\pi i}{3}}\left\vert 0101\right\rangle +e^{-%
\frac{2\pi i}{3}}\left\vert 0110\right\rangle \right.   \notag \\
&&+\left. e^{-\frac{2\pi i}{3}}\left\vert 1001\right\rangle +e^{\frac{2\pi i%
}{3}}\left\vert 1010\right\rangle +\left\vert 1100\right\rangle \right)
\end{eqnarray}%
is the four-qubit Dicke state [21] with two excitations if neglecting the
relative phases and $\left\vert \bar{\Psi}\right\rangle $ is the conjugate $%
\left\vert \Psi \right\rangle $. However, so far we can not find a proper
decomposition in terms of two-qubit projectors for $N_{i}$ (Strictly
speaking, we can not find such decompositions that reduce the number of
observables). That is to say, the projectors of $N_{i}$ have to be performed
on a four-qubit space as a whole. In this sense, only \textit{6} groups of
local projective measurements are enough for concurrence.

More interestingly, from eq. (3) one can find that $\tau $ (3-tangle) for a
tripartite pure state of qubits can be directly measured only by local
projective measurements on the subsystems of two qubits. Consider a
tripartite quantum pure state of qubits $\left\vert \Psi \right\rangle _{ABC}
$ with the reduced density $\rho _{AB}$, then $\rho _{AB}$ is obviously
rank-two. Based on ref. [18,19], 3-tangle of $\left\vert \Psi \right\rangle
_{ABC}$ can be given by eq. (3). The above procedure implies eq. (3) is
measurable, i.e. 3-tangle is directly measurable if four copies of $\rho
_{AB}$ are available. Different from our previous work [20] that requires
projective measurements on all subsystems, only projective measurements on
two subsystems (reduced density matrix of two qubits) are enough, even
though so far projection measurements on four-qubit space are required.

We have expressed concurrence of rank-two mixed states by Hermite
operators which can be decomposed into local projectors. This shows
in principle that the exact concurrence can be directly and locally
measured in experiment. However, as mentioned above, it requires
projective measurements on the whole space of four qubits, which
means the interactions of four qubits. Take a photon experiment as
an example, it might need the interference of four photons. In the
current experiment, together with four copy of a state, it might be
very difficult to realize, hence present scheme might introduce no
advantage in current experiment. But we can safely say that only two
qubit interaction can not be enough for the really applicable
quantum computer. In this sense, the current work should be a
valuable contribution for the further research. In addition, it is
still an open problem whether the observables given here are derived
from an optimal decomposition such that the number of observables
are minimal and whether there exist some other decompositions of
$N_{i}$ or $\mathcal{A}$ which lead to a direct contribution to the
current experiments. This is our forthcoming efforts.

\section{Conclusion and discussion}

We have shown that the exact bipartite concurrence for rank 2 mixed states
of qubits instead of a lower bound is directly and locally measurable in
experiment, provided that four copies of the states of interests are
available. In particular, it is very interesting that 3-tangle for a
tripartite quantum pure state of qubits has been shown to be measurable
experimentally if only the \emph{fourfold} copy of bipartite reduced density
matrix is available, which is different from our previous work [20].
Furthermore, because a state has to be prepared repeatedly in order to
obtain reliable measurement statistics in any experiment [2], a fourfold
copy of a state should be feasible in principle, which implies the
observation of mixed-state entanglement may be feasible. However, practical
experiments may lead to four different "copies" of a state and influence the
fidelity, which requires that one has to perform necessary error analysis
based on different experimental realization [2].

\section{Acknowledgement.}

This work was supported by the National Natural Science Foundation
of China, under Grant Nos. 10747112, 10575017 and 60703100.

\end{document}